# Tunable Hybrid Qubit in a GaAs Double Quantum Dot

Gang Cao,[1, 2, *] Hai-Ou Li,[1, 2, *] Guo-Dong Yu,[1, 2] Bao-Chuan Wang ,[1, 2] Bao-Bao Chen,[1, 2] Xiang-Xiang Song,[1, 2] Ming Xiao,[1, 2, †] Guang-Can Guo,[1, 2] Hong-Wen Jiang,[3] Xuedong Hu,[4] and Guo-Ping Guo[1, 2, †]

1 Key Laboratory of Quantum Information, University of Science and Technology of China,

Chinese Academy of Sciences, Hefei 230026, China

2 Synergetic Innovation Center of Quantum Information & Quantum Physics,

University of Science and Technology of China, Hefei, Anhui 230026, China

3 Department of Physics and Astronomy, University of California at Los Angeles,

California 90095, USA

4 Department of Physics, University at Buffalo, SUNY, Buffalo, New York 14260, USA

*These authors contributed equally to this work

†Corresponding Authors: gpguo@ustc.edu.cn; maaxiao@ustc.edu.cn

# Abstract

We experimentally demonstrate a tunable hybrid qubit in a five-electron GaAs double quantum dot. The qubit is encoded in the (1,4) charge regime of the double dot, and can be manipulated completely electrically. More importantly, dot anharmonicity leads to quasi-parallel energy levels and a new anti-crossing, which help preserve quantum coherence of the qubit and yield a useful working point. We have performed Larmor precession and Ramsey fringe experiments near the new working point, and find that the qubit decoherence time is significantly improved over a charge qubit. This work shows a new way to encode a semiconductor qubit that is controllable and coherent.

The promise of massive computing power from the intrinsic parallelism in a quantum system has driven extensive research activities on spin and charge dynamics in semiconductor nanostructures [1-3]. Over the past decade, significant experimental progress has been achieved in demonstrating coherent manipulation of spin- and charge-based qubits. Specifically, a variety of spin qubits have been shown to have extremely long coherence times, though high-fidelity two-qubit gates remain a difficult technical challenge [4-12]. Charge qubits [13-21], on the other hand, can be manipulated very fast because of the strong electrical interaction, which unfortunately also leads to usually short coherence times. Finding a balance between coherence and maneuverability is thus still an open problem in the pursuit of a scalable solid state quantum computer.

Hybrid qubit [22, 23] in a Si/SiGe heterostructure is one of the proposals that attempt to address this issue [24-26]. By encoding into two three-electron states in a double dot that have different spin symmetries, relaxation is suppressed. The qubit states also have extended sweet spots, which reduce the effect of charge-noise-induced dephasing [27-33]. Lastly, the qubit can be manipulated purely electrically. Recent experiments in SiGe-based devices have clearly shown that hybrid qubits are indeed really fast, and their coherence times are significantly longer than charge qubits [34, 35].

An intriguing question for us is whether the hybrid qubit design can be adapted to a GaAs double quantum dot. While as a host material for spin qubits GaAs suffers significantly from all the nuclear spins in the lattice, the fast operation speed of a hybrid qubit may alleviate this problem. The lighter electron mass and simpler band structure in GaAs should definitely help in terms of qubit tunability, as has been shown repeatedly in the past quantum dot research work in GaAs [2].

Here we experimentally demonstrate a single tunable hybrid qubit in a GaAs double quantum dot. By working near the (2,3)-(1,4) instead of (2,1)-(1,2) charge transition, we succeed in tuning our hybrid qubit to an energy splitting in the microwave frequency range. More importantly, asymmetry and anharmonicity in one of the dots leads to a new anti-crossing point between the quasi-parallel energy levels of the qubit, yielding a versatile and coherent working point. We have performed both Larmor precession and Ramsey fringe experiments on this qubit, and demonstrated that at this new working point the qubit has a $T_2^*$ time in the order of 10 ns, much longer than a normal charge qubit. Our results clearly illustrated how a few-electron system can have high degrees of tunability and quantum coherence, and appropriate frequency for qubit control. The simple band structure of GaAs allows us to establish a clear understanding on how our qubit works. Further explorations in a GaAs double dot could also help us clarify the roles played by the various electrical noises such as piezoelectric and polar couplings to phonons. Furthermore, the general understanding of the roles played by level degeneracy and dot deformation can also lead to better qubit designs in other (potentially more coherent) substrate materials such as isotopically purified Si.

The system we study is a conventional gated double dot on a $GaAs/Al_{0.3}Ga_{0.7}As$ heterostructure. Figure 1(a) is a scanning electron microscopy (SEM) image of the device. Together with H1 and H2 gates, D1 through D5 gates define the double dot, and Q1 through Q3 gates define a quantum point contact (QPC). The QPC works at a 0.2 mV dc bias, and acts as a charge sensor for the qubit. With D5 gate determining the energy detuning across the double dot, a particularly interesting quantity is the QPC's transconductance $dI/dV_{D5}$, which is obtained using a lock-in amplifier and superposing a small ac voltage of amplitude of 0.1 mV on D5. Microwave signals can also be applied on gate D5, through a semi-rigid coaxial transmission line from room temperature. We are able to access the few-electron regime for this double dot, as illustrated by the charge stability diagram given in Fig. 1(b).

Denoting the electron number in the left and right dots as (N,M), we focus on the low-energy electron dynamics in the region near the (2,3)-(1,4) charge transition for the rest of this paper. This region is not exactly where the original hybrid qubit works, which was designed near the (2,1)-(1,2) transition. However, since the two core electrons in the right dot occupying the ground orbital state do not participate in the low-energy dynamics at the crossing, the relevant physics is very similar to the three-electron physics between (2,1) and (1,2) configurations. Thus we will continue to use the three-electron nomenclature to describe the dynamics near the (2,3)-(1,4) crossing [36].

The low-energy sector of the double dot spectrum consists of three five-electron states. Suppose the lowest-energy orbital states in the right dot are $\psi_0$, $\psi_1$ and $\psi_2$. With $\psi_0$ occupied by the two core electrons, the additional one electron in the right dot in the (2,3) regime will occupy $\psi_1$, while the additional two electrons in the (1,4) regime will both occupy $\psi_1$ if they form a singlet, or $\psi_1$ and $\psi_2$ if they form a triplet state. Thus near the charge transition there are three relevant low-energy states [36]: $|1\rangle = |S\rangle_L |\uparrow\rangle_R$, $|2\rangle = |\uparrow\rangle_L |S\rangle_R$, and $|3\rangle = \frac{1}{\sqrt{3}}|\uparrow\rangle_L |T_0\rangle_R - \frac{2}{\sqrt{3}}|\downarrow\rangle_L |T_\uparrow\rangle_R$, in the $S_z = \frac{1}{2}$ manifold (the $S_z = -\frac{1}{2}$ manifold is degenerate with and decoupled from the $S_z = \frac{1}{2}$ manifold, and has identical charge dynamics. However, in the absence of an applied magnetic field, the nuclear spins in the substrate could cause these two manifolds to mix and limits our $T_2^*$ time), identical to those for a hybrid qubit. States $|2\rangle$ and $|3\rangle$ form our hybrid qubit in the (1,4) region, where they are the two lowest-energy states.

Our explorations of charge transitions indicate that different orbital states in the right dot have different charge responses to the applied gate voltages. Figure 2(a) shows the charge stability diagram near the (2,3)-(1,4) transition. Here an additional short pulse of duration $T_p = 300$ ps and magnitude $V_p = 135$ μeV is repeatedly superposed onto the DC sweep of the D5 gate [17, 18]. Two different sets of charge oscillation peaks can be observed, marked by the blue and purple dashed lines. They clearly have distinct slopes, which indicate that orbital states $\psi_1$ and $\psi_2$ (which are the underlying orbital states for states $|2\rangle$ and $|3\rangle$) have different lever arms relative to D3 and D5 gates. Consequently, when the gate voltage on the right dot is varied, the energies of states $|2\rangle$ and $|3\rangle$ change differently. In fact, this variation in the response to the applied gate voltages can already be seen in Fig. 1(b), where two different charging lines with slopes -1.68 and -2.17 are marked with purple and blue dashed lines.

The different responses by $\psi_1$ and $\psi_2$ to the applied voltages can be explained by the two states having different 'centers of charge', which in turn indicates that the right dot is strongly anharmonic. In a harmonic dot, $\psi_1$ and $\psi_2$ would have corresponded to the two degenerate P orbitals of the Fock-Darwin states that are symmetric around the dot center, and would have responded to the applied gate voltages in the same manner at the lowest order. In our sample, on the other hand, the double dot energy diagram is shown in Fig. 2(b), where states $|2\rangle$ and $|3\rangle$ are not parallel as functions of the energy detuning. Instead they cross at a particular detuning, and the crossing becomes an anti-crossing after tunnel coupling to the left dot is taken into consideration. Compared with the original hybrid qubit, the Hamiltonian of this three-level system can be written as:

$$H = \begin{pmatrix} \varepsilon/2 & \Delta_1 & \Delta_2 \\ \Delta_1 & -\varepsilon/2 & 0 \\ \Delta_2 & 0 & -k\varepsilon/2 + \delta \end{pmatrix} \quad (1)$$

Here $\varepsilon$ is the energy detuning between the three states, and can be tuned across the inter-dot charge transition (as shown in Fig. 2a), $\delta$ is the singlet-triplet splitting within the right dot, and $k$ represents the different lever arms of states $|2\rangle$ and $|3\rangle$. Among the off-diagonal terms, and are tunnel couplings from $|1\rangle$ to $|2\rangle$ and $|3\rangle$. Direct coupling between $|2\rangle$ and $|3\rangle$ vanishes because of their different spin symmetries.

Instead of having a constant singlet-triplet splitting, states $|2\rangle$ and $|3\rangle$ now have different dependences on the gate voltage, represented by the new phenomenological parameter $k$ for state $|3\rangle$. Since there are two different sets of transition lines mixed with each other in Fig. 2(a), states $|2\rangle$ and $|3\rangle$ should be nearly degenerate at $\varepsilon \sim 0$, which means $\delta \sim 0$. Thus for simplicity we choose $\delta = 0$ in our calculations. Figure 2(b) shows the energy levels calculated using Eq. (1) with $2\Delta_1 = 9.2$ GHz, $2\Delta_2 = 7.5$ GHz, $k = 1.3$ and $\delta = 0$, while Fig. 2(c) gives the energy splitting between the two lowest levels. States $|2\rangle$ and $|3\rangle$ are not parallel anymore, with level spacing reaching a minimum of ~ 2.5 GHz at $\varepsilon \sim 55\,\mu\text{eV}$.

To verify our model, we characterize the energy levels using microwave spectroscopy [21]. For this and following experiments, a lock-in amplifier's TTL signal is used to chop the pulses to generate the sequences, so that half of the TTL period has no pulse, and the lock-in amplifier is also used to measure the QPC's corresponding current change dI [35].

As shown schematically in Fig. 2(b), we initialize our double dot at a negative detuning point, where the ground state, state $|1\rangle$, is non-degenerate and well separated from the excited states. Consequently our initialization procedure is quite robust. We then increase the detuning adiabatically, over a time period of 10 ns, to the region around the $|2\rangle$-$|3\rangle$ anti-crossing. A 20 ns microwave pulse is then applied. Only on resonance can the qubit be excited from the ground state to the excited state. The system is then measured after the detuning is adiabatically lowered in another 10 ns. Figure 2(d) shows the resulting QPC conductance signal. The resonant line agrees well with the calculated result from Hamiltonian (1) presented in Fig. 2(c).

Figure 2 shows that, compared with the original hybrid qubit [19, 21], our qubit has some distinct characteristics. Because of the anharmonicity and asymmetry in the right dot, our single-dot energy levels can be modified by changing the gate voltage, making this hybrid qubit tunable. On the negative side, the almost parallel dispersion of the two hybrid qubit states in the original design (and in Fig. S1 of [36]) becomes a non-parallel dispersion when we adjust D3 gate voltage, which potentially weakens the qubit against decoherence effects of the environmental charge noise. Considering that hybrid qubit was designed to suppress effects of charge noise by employing parallel energy levels [35, 37], it is crucial to clarify the coherence properties of our variant. Below we explore how the changes of our qubit, particularly the non-parallel energy levels, affect its coherence.

We measure Larmor oscillations and Ramsey fringes near the working point of our hybrid qubit in order to characterize its decoherence. As in the microwave spectroscopy experiment, we initialize the qubit in the ground state $|1\rangle$ in the (2,3) region at $\varepsilon \sim -55\,\mu\text{eV}$. We then take the qubit to $\varepsilon \sim 0$ adiabatically using a pulse with a rise time of 3 ns, as shown in Fig. 2(b). The system remains in the ground state. We then apply a non-adiabatic pulse to drive the qubit to $\varepsilon \sim 55\,\mu\text{eV}$, deep in the (1,4) region, where the qubit will evolve freely, like the Larmor precession of a spin in a magnetic field, with the excited state acquiring an additional phase relative to the ground state that is proportional to the pulse width $T_p$. After this free evolution at $\varepsilon \sim 55\,\mu\text{eV}$, the detuning is returned non-adiabatically to $\varepsilon \sim 0$, and then adiabatically to $\varepsilon \sim -55\,\mu\text{eV}$ in 2 ns, at which point we measure the QPC current. If the qubit has evolved to the ground (first excited) state at $\varepsilon \sim 55\,\mu\text{eV}$, it would be in $|1\rangle$ ($|2\rangle$) at the end of the pulse sequence. Since states $|1\rangle$ and $|2\rangle$ have different numbers of electrons in the left dot, they can be differentiated by the QPC.

Figure 3(a) shows the results of Larmor precession of the qubit as a function of the pulse width $T_p$ at detuning $\varepsilon \sim 55\,\mu\text{eV}$, or where $\varepsilon_t \sim 0$ ($\varepsilon_t$ is defined as the energy detuning relative to the new anti-crossing point between the qubit levels). The minimum energy gap at $\varepsilon_t = 0$ is obtained from the oscillation frequency of $2\Delta = 2.43\,\text{GHz}$, which is consistent with the minimum frequency we extract from Fig. 2(d) using microwave spectroscopy. Figure 3(b) is a cut along the green dashed line in Fig. 3(a), at $\varepsilon_t \sim 0$. Fitting the curve using $A\exp\left(-(t-t_0)^2/T_2^{*2}\right)\cos(\omega t + \theta)$, we obtain $T_2^* = 8.10\,\text{ns}$. This decoherence time is much longer than the value we obtained in the charge qubit regime [36], and values obtained in other charge qubit experiments [13, 19, 20].

This significant increase in decoherence time originates from the anti-crossing of the quasi-parallel energy levels in Fig. 2(c). Specifically, the anti-crossing and the corresponding optimal point suppresses the charge noise induced dephasing. Furthermore, the different spin symmetries of the qubit states suppress the qubit relaxation. Therefore, even though our qubit has lost some of the coherence preserving ability of the original hybrid qubit design, it still possesses much better coherence properties than an ordinary charge qubit. Moreover, compared with the original hybrid qubit, the new anti-crossing point of this qubit, which resides completely in the (1,4) charge region, enables *x*-axis rotations on the Bloch sphere with a long coherence time.

We have also performed a Ramsey-fringe experiment to further clarify the coherence properties of our qubit in the vicinity of the anti-crossing point at $\varepsilon_t \sim 0$. Figure 4(a) presents the pulse shape of the Ramsey experiment, which consists of a free evolution around the z-axis in between two $3\pi/2$ rotations around the x-axis. Varying the amplitude $V_t$ of the middle pulse between the two x-rotations allows the qubit to reach different detuning points.

Figure 4(b) shows the Ramsey fringes as a function of the free evolution point $\varepsilon_t$ and the evolution duration $T_w$. As expected, the Ramsey fringes decay rapidly, in sub-nanoseconds, when $\varepsilon_t < 0$ and the qubit is in the charge-qubit regime, near the (2,3)-(1,4) charge transition. The different charge distributions of the qubit states lead to fast decoherence due to electrical noises [13, 17]. In contrast, when $\varepsilon_t \geq 0$, the qubit is deep in the (1,4) charge region, with both qubit states having similar charge distribution. The slowly increasing energy gap also provides some protection for the qubit against charge noise. As a result, the fringes remain quite visible even beyond 6 ns, and the qubit has much longer decoherence time than in the $\varepsilon_t < 0$ region. The remaining decoherence is possibly due to hyperfine interaction with nuclear spins and residue effects of the charge noise because our levels are only quasi-parallel [38]. Figure 4(c) shows three representative cuts, at $\varepsilon_t = 30$, 0 and $-30\,\mu\text{eV}$ from top to bottom respectively. The corresponding decoherence time $T_2^*$ extracted from the curves are $T_2^* \sim$ 5.8, 8.30 and 0.74 ns. Figure 4(d) presents decoherence time $T_2^*$ obtained from Figure 4(b), as a function of the detuning point $\varepsilon_t$.

We note here that the presence of the new anti-crossing point for our hybrid encoding allows qubit operations around both x- and z-axis with an enhanced coherence time $T_2^*$ of up to 8.3 ns. This is different from the Si hybrid qubit, where x- and z-rotations are performed at different inter-dot detuning, so that they have different coherence times [35]. Furthermore, on the positive detuning side for our system [deeper into the (1,4) charge regime], $T_2^*$ only decreases gradually and slowly due to the quasi-parallel energy levels.

In conclusion, we have experimentally demonstrated a tunable hybrid qubit in a five-electron GaAs double quantum dot. Specifically, we show that the qubit is tunable because of an anharmonic and asymmetric quantum dot potential. The qubit energy levels anti-cross instead of being parallel, yielding a new working point. The non-parallel energy levels make the qubit more susceptible to charge noise, but the new working point gives it higher degree of tunability. To clarify the controllability and coherence of this new qubit, we have performed Larmor precession and Ramsey fringe experiments. We find the qubit decoherence time is quite long, in the order of 10 ns, when it is deep in the (1,4) charge configuration, much slower than in the regular charge qubit regime.

Our results show that this variant of hybrid qubit in a GaAs double quantum dot is both controllable and coherent, and acts as a useful platform to study coherent control in a semiconductor nanostructure. To establish its viability as a practical scalable qubit, more work is needed in establishing high fidelity gates, and to identify the optimal approach for a two-qubit gate. In addition, splitting our right dot into a double dot could make it even more tunable, and could open up new avenues for qubit manipulation.

**Acknowledgements:** This work was supported by the NFRP (Grant No. 2011CBA00200), the SPRP of the CAS (Grant No. XDB01030000), and the NNSF (Grants No. 11304301, 11575172, 61306150 and 91421303). XH and HWJ acknowledge financial support by US ARO through grant W911NF0910393 and W911NF1410346, respectively.

**FIGURE CAPTIONS**

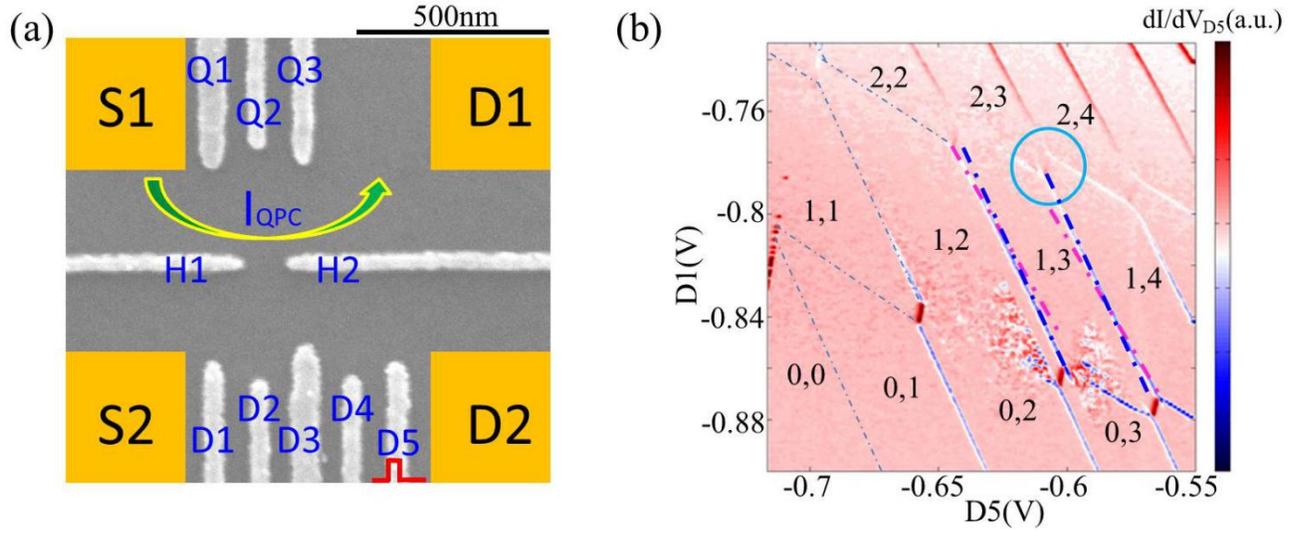

**FIG 1.** (a) A SEM image of the sample structure. The density of the 2DEG (95nm below surface) is $3.2 \times 10^{11}$ cm$^{-2}$ with a mobility of $1.5 \times 10^{5}$ cm$^{2}/$Vs. The electron temperature in the device is about 150 mK. (b) Charge diagram of the double quantum dot at $D3 = -0.25$ V, $D4 = -0.35$ V, $H2 = -0.7$ V. The voltages of gates D2 and H1 are kept as constant, too. The purple and blue dashed lines present two different charging lines with slop of -1.68, -2.17 respectively. To make it clearly visible, the two dashed lines are extended along the charging lines.

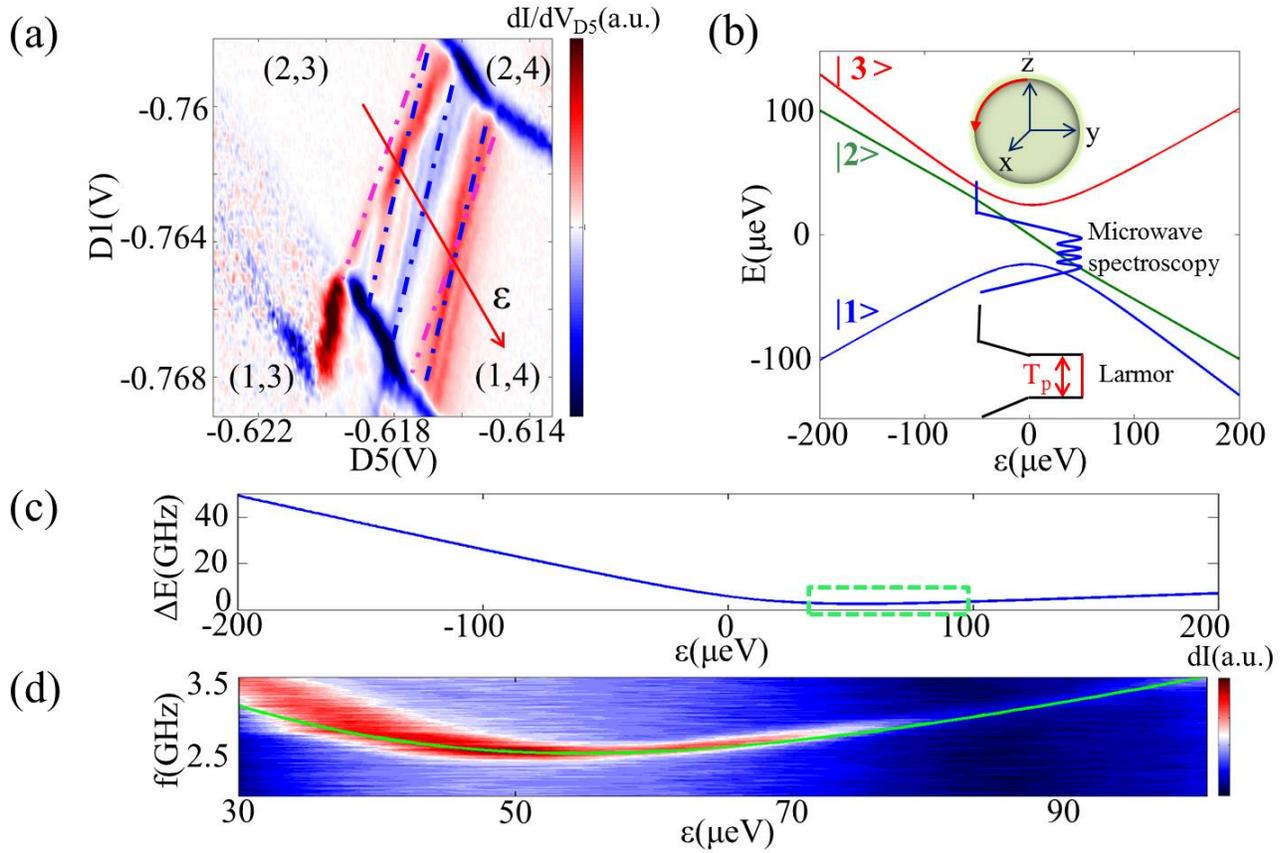

**FIG 2.** (a) Charge diagrams under a short pulse $T_p = 300 \, \text{ps}$, $V_p = 135 \, \mu\text{eV}$ at gate voltages of $D3 = -0.217 \, \text{V}$, $D4 = -0.417 \, \text{V}$ and $H2 = -0.717 \, \text{V}$. (b) The simulated energy levels of the system. (c) The energy gap between the ground state and the first excited state in (b). (d) QPC signal as a function of the excitation frequency f of the microwave and the detuning energy. The green line is the energy gap given in the dashed box of (c).

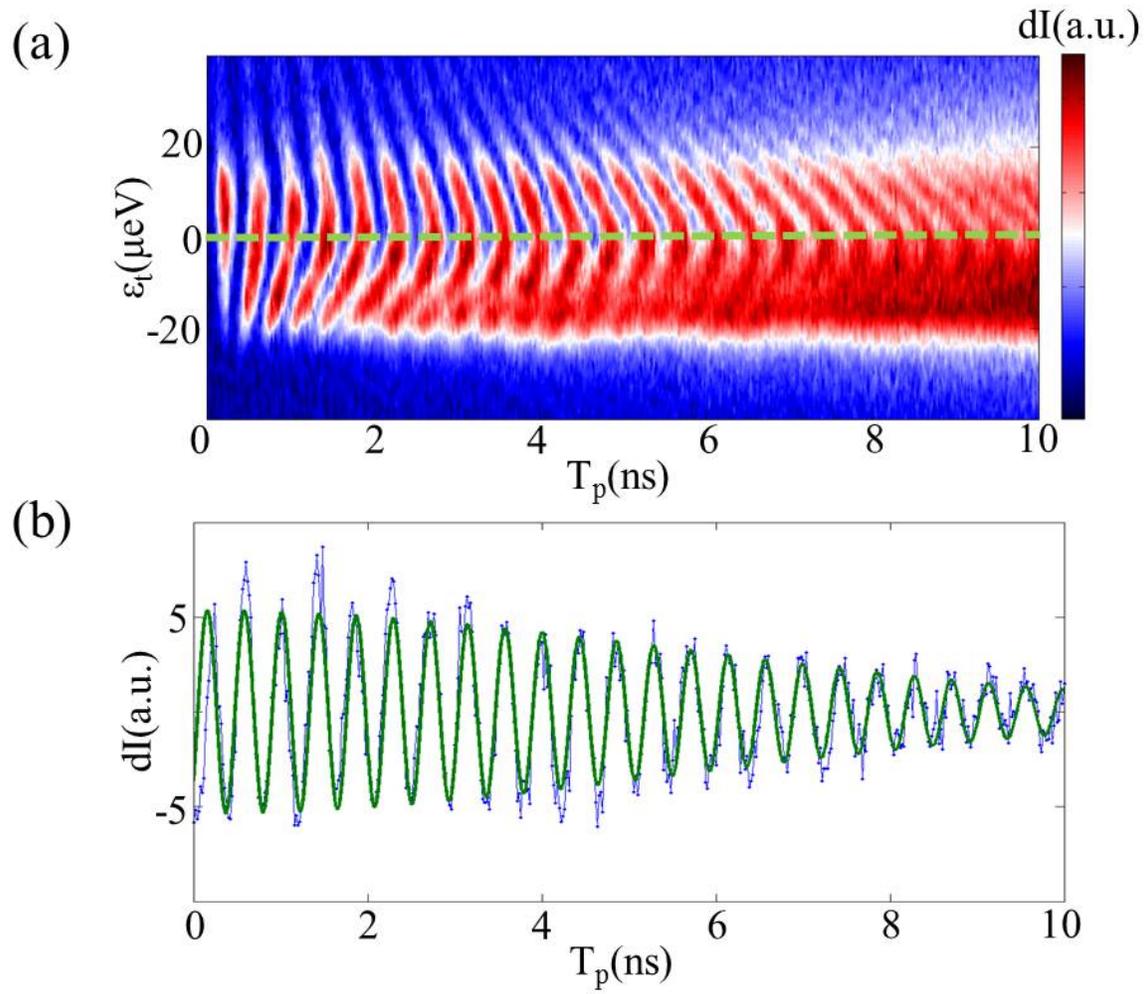

**FIG 3.** (a) The Larmor precession and the associated charge oscillations as a function of the non-adiabatic pulse width $T_p$ and $\varepsilon_t$. The signal along the green dashed line is plotted in (b) with blue, while the green solid line is a numerical fit after subtraction of the background, and yields a dephasing time of $T_2^* = 8.10$ ns.

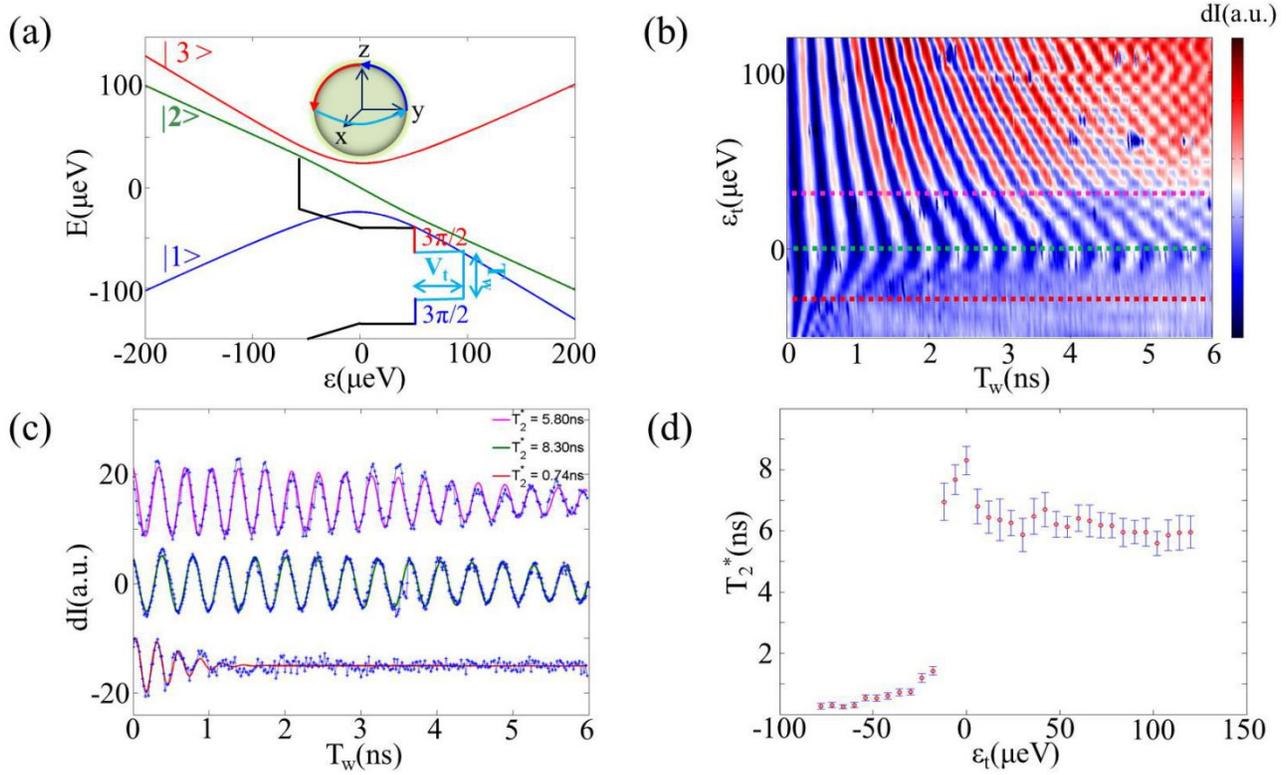

**FIG 4.** (a) The simulated energy levels and the pulse shape of Ramsey fringes. The trajectory segments with different colors on the Bloch sphere correspond to the different parts of the applied pulse with the same colors. (b) Ramsey fringes as a function of the inserted pulse width $T_w$ and amplitude $\varepsilon_t$. After subtraction of the background signal, the three dashed lines at $\varepsilon_t = 30$, $0$ and $-30$ μeV are presented and fitted from top to bottom in (c) with an offset current. (d) The extracted $T_2^*$ as a function of detuning energy $\varepsilon_t$.

# Supplementary Material: Tunable Hybrid Qubit in a GaAs Double Quantum Dot

Gang Cao,[1,2,*] Hai-Ou Li,[1,2,*] Guo-Dong Yu,[1,2] Bao-Chuan Wang,[1,2] Bao-Bao Chen,[1,2] Xiang-Xiang Song,[1,2] Ming Xiao,[1,2,†] Guang-Can Guo,[1,2] Hong-Wen Jiang,[3] Xuedong Hu,[4] and Guo-Ping Guo[1,2,†]

[1] Key Laboratory of Quantum Information, University of Science and Technology of China,
Chinese Academy of Sciences, Hefei 230026, China
[2] Synergetic Innovation Center of Quantum Information & Quantum Physics,
University of Science and Technology of China, Hefei, Anhui 230026, China
[3] Department of Physics and Astronomy, University of California at Los Angeles,
California 90095, USA
[4] Department of Physics, University at Buffalo, SUNY, Buffalo, New York 14260, USA

## I. THEORETICAL MODEL OF A FIVE-ELECTRON DOUBLE DOT

As discussed in the main text, the working point of our qubit is near the $(2, 3) - (1, 4)$ charge transition. Neglecting contributions from higher-energy orbitals and inter-dot overlaps via configuration interaction, the lowest-energy states can be represented by five-electron Slater determinants:

$$|1\rangle = \frac{1}{\sqrt{5!}} \begin{vmatrix} \psi_{L\uparrow}(1) & \psi_{L\uparrow}(2) & \psi_{L\uparrow}(3) & \psi_{L\uparrow}(4) & \psi_{L\uparrow}(5) \\ \psi_{L\downarrow}(1) & \psi_{L\downarrow}(2) & \psi_{L\downarrow}(3) & \psi_{L\downarrow}(4) & \psi_{L\downarrow}(5) \\ \psi_{0\uparrow}(1) & \psi_{0\uparrow}(2) & \psi_{0\uparrow}(3) & \psi_{0\uparrow}(4) & \psi_{0\uparrow}(5) \\ \psi_{0\downarrow}(1) & \psi_{0\downarrow}(2) & \psi_{0\downarrow}(3) & \psi_{0\downarrow}(4) & \psi_{0\downarrow}(5) \\ \psi_{1\uparrow}(1) & \psi_{1\uparrow}(2) & \psi_{1\uparrow}(3) & \psi_{1\uparrow}(4) & \psi_{1\uparrow}(5) \end{vmatrix},$$

$$|2\rangle = \frac{1}{\sqrt{5!}} \begin{vmatrix} \psi_{L\uparrow}(1) & \psi_{L\uparrow}(2) & \psi_{L\uparrow}(3) & \psi_{L\uparrow}(4) & \psi_{L\uparrow}(5) \\ \psi_{0\uparrow}(1) & \psi_{0\uparrow}(2) & \psi_{0\uparrow}(3) & \psi_{0\uparrow}(4) & \psi_{0\uparrow}(5) \\ \psi_{0\downarrow}(1) & \psi_{0\downarrow}(2) & \psi_{0\downarrow}(3) & \psi_{0\downarrow}(4) & \psi_{0\downarrow}(5) \\ \psi_{1\uparrow}(1) & \psi_{1\uparrow}(2) & \psi_{1\uparrow}(3) & \psi_{1\uparrow}(4) & \psi_{1\uparrow}(5) \\ \psi_{1\downarrow}(1) & \psi_{1\downarrow}(2) & \psi_{1\downarrow}(3) & \psi_{1\downarrow}(4) & \psi_{1\downarrow}(5) \end{vmatrix},$$

$$|4\rangle = \frac{1}{\sqrt{2\cdot 5!}} \begin{vmatrix} \psi_{L\uparrow}(1) & \psi_{L\uparrow}(2) & \psi_{L\uparrow}(3) & \psi_{L\uparrow}(4) & \psi_{L\uparrow}(5) \\ \psi_{0\uparrow}(1) & \psi_{0\uparrow}(2) & \psi_{0\uparrow}(3) & \psi_{0\uparrow}(4) & \psi_{0\uparrow}(5) \\ \psi_{0\downarrow}(1) & \psi_{0\downarrow}(2) & \psi_{0\downarrow}(3) & \psi_{0\downarrow}(4) & \psi_{0\downarrow}(5) \\ \psi_{1\uparrow}(1) & \psi_{1\uparrow}(2) & \psi_{1\uparrow}(3) & \psi_{1\uparrow}(4) & \psi_{1\uparrow}(5) \\ \psi_{2\downarrow}(1) & \psi_{2\downarrow}(2) & \psi_{2\downarrow}(3) & \psi_{2\downarrow}(4) & \psi_{2\downarrow}(5) \end{vmatrix}$$

$$+ \frac{1}{\sqrt{2\cdot 5!}} \begin{vmatrix} \psi_{L\uparrow}(1) & \psi_{L\uparrow}(2) & \psi_{L\uparrow}(3) & \psi_{L\uparrow}(4) & \psi_{L\uparrow}(5) \\ \psi_{0\uparrow}(1) & \psi_{0\uparrow}(2) & \psi_{0\uparrow}(3) & \psi_{0\uparrow}(4) & \psi_{0\uparrow}(5) \\ \psi_{0\downarrow}(1) & \psi_{0\downarrow}(2) & \psi_{0\downarrow}(3) & \psi_{0\downarrow}(4) & \psi_{0\downarrow}(5) \\ \psi_{1\downarrow}(1) & \psi_{1\downarrow}(2) & \psi_{1\downarrow}(3) & \psi_{1\downarrow}(4) & \psi_{1\downarrow}(5) \\ \psi_{2\uparrow}(1) & \psi_{2\uparrow}(2) & \psi_{2\uparrow}(3) & \psi_{2\uparrow}(4) & \psi_{2\uparrow}(5) \end{vmatrix},$$

$$|5\rangle = \frac{1}{\sqrt{5!}} \begin{vmatrix} \psi_{L\downarrow}(1) & \psi_{L\downarrow}(2) & \psi_{L\downarrow}(3) & \psi_{L\downarrow}(4) & \psi_{L\downarrow}(5) \\ \psi_{0\uparrow}(1) & \psi_{0\uparrow}(2) & \psi_{0\uparrow}(3) & \psi_{0\uparrow}(4) & \psi_{0\uparrow}(5) \\ \psi_{0\downarrow}(1) & \psi_{0\downarrow}(2) & \psi_{0\downarrow}(3) & \psi_{0\downarrow}(4) & \psi_{0\downarrow}(5) \\ \psi_{1\uparrow}(1) & \psi_{1\uparrow}(2) & \psi_{1\uparrow}(3) & \psi_{1\uparrow}(4) & \psi_{1\uparrow}(5) \\ \psi_{2\uparrow}(1) & \psi_{2\uparrow}(2) & \psi_{2\uparrow}(3) & \psi_{2\uparrow}(4) & \psi_{2\uparrow}(5) \end{vmatrix},$$

$$|3\rangle = \frac{1}{\sqrt{3}}|4\rangle - \frac{2}{\sqrt{3}}|5\rangle,$$

where the numbers in the parenthesis are labels for the five electrons, $\psi_L$ is the single-electron ground orbital state in the left dot, and $\psi_i$ with $i=1,2,3$ are the three lowest-energy single-electron orbitals in the right dot. The two core electrons in the right dot occupy the lowest orbital state $\psi_0$ for all three states, and do not participate in the low-energy dynamics of this double dot. Their presence only results in energy shifts through various Coulomb matrix elements, and is not affected by inter-dot detuning. Thus we can neglect them when considering qubit dynamics from changing the inter-dot detuning energy, and denote the five states above in the same nomenclature as those for the three-electron Si hybrid qubit: $|1\rangle = |S\rangle_L|\uparrow\rangle_R$, $|2\rangle = |\uparrow\rangle_L|S\rangle_R$, $|4\rangle = |\uparrow\rangle_L|T_0\rangle_R$, $|5\rangle = |\downarrow\rangle_L|T_\uparrow\rangle_R$, and $|3\rangle = \frac{1}{\sqrt{3}}|\uparrow\rangle_L|T_0\rangle_R - \frac{2}{\sqrt{3}}|\downarrow\rangle_L|T_\uparrow\rangle_R$. Here $|\uparrow\rangle_L = \psi_L(\vec{r})|\uparrow\rangle$, $|\uparrow\rangle_R = \psi_1(\vec{r})|\uparrow\rangle$, $|S\rangle_L = \psi_L(\vec{r}_1)\psi_L(\vec{r}_2) \times \frac{1}{\sqrt{2}}|\uparrow\downarrow - \downarrow\uparrow\rangle$, $|S\rangle_R = \psi_1(\vec{r}_1)\psi_1(\vec{r}_2) \times \frac{1}{\sqrt{2}}|\uparrow\downarrow - \downarrow\uparrow\rangle$,

$$|T_0\rangle_R = \frac{1}{\sqrt{2}}[\psi_1(\vec{r}_1)\psi_2(\vec{r}_2) - \psi_2(\vec{r}_1)\psi_1(\vec{r}_2)] \times \frac{1}{\sqrt{2}}|\uparrow\downarrow + \downarrow\uparrow\rangle \quad , \quad \text{and}$$

$$|T_\uparrow\rangle_R = \frac{1}{\sqrt{2}}[\psi_1(\vec{r}_1)\psi_2(\vec{r}_2) - \psi_2(\vec{r}_1)\psi_1(\vec{r}_2)] \times |\uparrow\uparrow\rangle \,.$$ Notice that without interdot tunneling, $|4\rangle$ and $|5\rangle$ are degenerate (even in the presence of a uniform magnetic field, since both belong to the $S_z = +\frac{1}{2}$ Zeeman sub-level). In the presence of interdot tunneling, the two are mixed, with one of the new eigenstates having lower energy and becoming $|3\rangle = \frac{1}{\sqrt{3}}|\uparrow\rangle_L|T_0\rangle_R - \frac{\sqrt{2}}{\sqrt{3}}|\downarrow\rangle_L|T_\uparrow\rangle_R$, while the other having higher energy and taking on the form of $|6\rangle = \frac{\sqrt{2}}{\sqrt{3}}|\uparrow\rangle_L|T_0\rangle_R + \frac{1}{\sqrt{3}}|\downarrow\rangle_L|T_\uparrow\rangle_R$. In our experiment we did not observe any evidence of this state, thus it is not included in our model. Our description of the low-energy dynamics of the double dot thus only involves three states: $|1\rangle$, $|2\rangle$, and $|3\rangle$. These are the basis states for Eq. (1) in the main text, and the latter two states form our hybrid qubit in the (1,4) charging regime.

In a Si/SiGe hybrid qubit, the energy difference between states $|2\rangle$ and $|3\rangle$ comes from the valley splitting. In other words $\psi_2$ is in a different valley eigenstate from that for $\psi_1$. Valley splittings in the SiGe heterostructures are generally in the vicinity of tens of μeV, which correspond to a microwave frequency in the order of 10 GHz. In GaAs, the singlet-triplet splitting in a two-electron single dot is normally in the range of meV, dominated by the orbital excitation energy [4]. However, here we have four electrons in the right dot. The two core electrons fill up the ground orbital state, so that the additional electrons in the right dot can only occupy excited states $\psi_1$ and/or $\psi_2$. In general, in the absence of a magnetic field, excited energy levels in a symmetric quantum dot always have degeneracies. For example, if the right dot here is circular and two-dimensional, $\psi_1$ and $\psi_2$ would be degenerate P orbitals of the Fock-Darwin state spectrum. Even if this degeneracy is lifted by the presence of asymmetry and/or magnetic field (of reasonable magnitude), the splitting would generally still be small, so that the total energy splitting between qubit states $|2\rangle$ and $|3\rangle$, which is dominated by the energy splitting between $\psi_1$ and $\psi_2$ but modified by Coulomb interaction, is also small (as compared to the singlet-triplet splitting in a two-electron GaAs quantum dot). In our case, the qubit energy splitting falls conveniently in the microwave frequency range.

## II.    DOUBLE DOT IN THE HYBRID QUBIT REGIME

With the core electrons ensuring a reasonable qubit energy splitting, our system behaves exactly like a Si/SiGe hybrid qubit when the right dot has a simple single dot spectrum. This is illustrated in Fig. S1. Panel (a) of Fig. S1 shows the charge stability diagram near the (2, 3) - (1, 4) charge transition, when a short pulse of duration $T_p = 300$ ps and magnitude $V_p = 200$ μeV is applied on the D5 gate in addition to the DC sweep across the charge transition [17-20], in order to generate Larmor precessions. Other gates are fixed at the voltages given in the figure caption. Compared to the charge stability diagram given in Fig. 1 of the main text, the major change here is in D3 voltage, which modifies the electric field over the right dot. The repeated peaks in QPC conductance indicate charge distribution changes due to transitions during the Larmor precessions.

In Fig. S1(b) we present a qualitative sketch of the energy diagram of $|1\rangle$, $|2\rangle$ and $|3\rangle$ near this charge transition. The corresponding Hamiltonian can be written as [34]:

$$H = \begin{pmatrix} \varepsilon/2 & \Delta_1 & \Delta_2 \\ \Delta_1 & -\varepsilon/2 & 0 \\ \Delta_2 & 0 & -\varepsilon/2 + \delta \end{pmatrix} \qquad (S1).$$

Here the diagonal terms indicate the charge distribution difference [(2, 3) to (1, 4)] between the three states, with $\delta$ the singlet-triplet splitting within the right dot. Among the off-diagonal terms, $\Delta_1$ and $\Delta_2$ are tunnel couplings from $|1\rangle$ to $|2\rangle$ and $|3\rangle$, respectively, while direct coupling between $|2\rangle$ and $|3\rangle$ vanishes because of their different spin symmetries. The Hamiltonian here is consistent with Eq. (1) in the main text with $k = 1$. In other words, the modification of the D3 voltage makes the charge centers of $\psi_1$ and $\psi_2$ coincide, so that they respond to the D5 voltage with roughly the same lever arm.

Hamiltonian (S1) is identical to that for a Si hybrid qubit [34], with the only significant difference in the origin of the singlet-triplet splitting as we have discussed above. Figure S1(c) shows the QPC conductance as a function of pulse length $T_p$ at different detuning points $\varepsilon_p$. The observation is similar to what was measured in a Si/SiGe hybrid qubit (such as Fig. 1 in Ref. [34]). The oscillations near the anti-crossing point between $|1\rangle$ and $|2\rangle$ are the typical Larmor precession for a charge qubit, or charge oscillations between the two dots [13, 17, 19, 20]. They are realized by pushing the system to the anti-crossing point where the true eigenstates are $(|1\rangle \pm |2\rangle)/\sqrt{2}$. The frequency of these oscillations is the coupling between $|1\rangle$ and $|2\rangle$, at $2\Delta_1 = 4.2\,\text{GHz}$, which is essentially the tunnel coupling between the two dots. In Fig. S1(d) we fit the Larmor oscillations at the $|1\rangle$-$|2\rangle$ anti-crossing and obtain a dephasing time of $T_2^* = 2.12\,\text{ns}$, which is consistent with earlier experimental results on charge qubits [13, 19, 20].

In the large positive detuning region of $\varepsilon_p > 50\,\mu\text{eV}$, additional coherent oscillations at a frequency of 12.0 GHz appear. They are more persistent than the Larmor oscillations at the $|1\rangle$-$|2\rangle$ anti-crossing, indicating a more robust coherent oscillation. These experimental observations can be well explained with Hamiltonian (S1) and Fig. S1(b) using $2\Delta_1 = 4.2\,\text{GHz}$, $2\Delta_2 = 5.0\,\text{GHz}$ and $\delta = 12.2\,\text{GHz}$. The high-detuning oscillations happen between states $|2\rangle$ and $|3\rangle$, which are the lowest-energy states deep in the (1, 4) regime (the positive-detuning regime) and have an energy splitting of 12.2 GHz. Since $\psi_1$ and $\psi_2$ have the same lever arm on the gates, states $|2\rangle$ and $|3\rangle$ would have the same dependence on the inter-dot detuning. They are therefore nearly parallel on Fig. S1(b), making the coherent oscillation between them more robust (with decoherence times as long as 7.7 ns) than the Larmor precession at the $|1\rangle$-$|2\rangle$ anti-crossing, comparable to the results for hybrid qubit in Si/SiGe [34, 35]. The extracted coherence time here is shorter than what we obtained in the main text, probably because the energy levels here are quasi-parallel but not exactly parallel, while in the main text the largest decoherence time is obtained at the anti-crossing, when the two levels are exactly parallel.

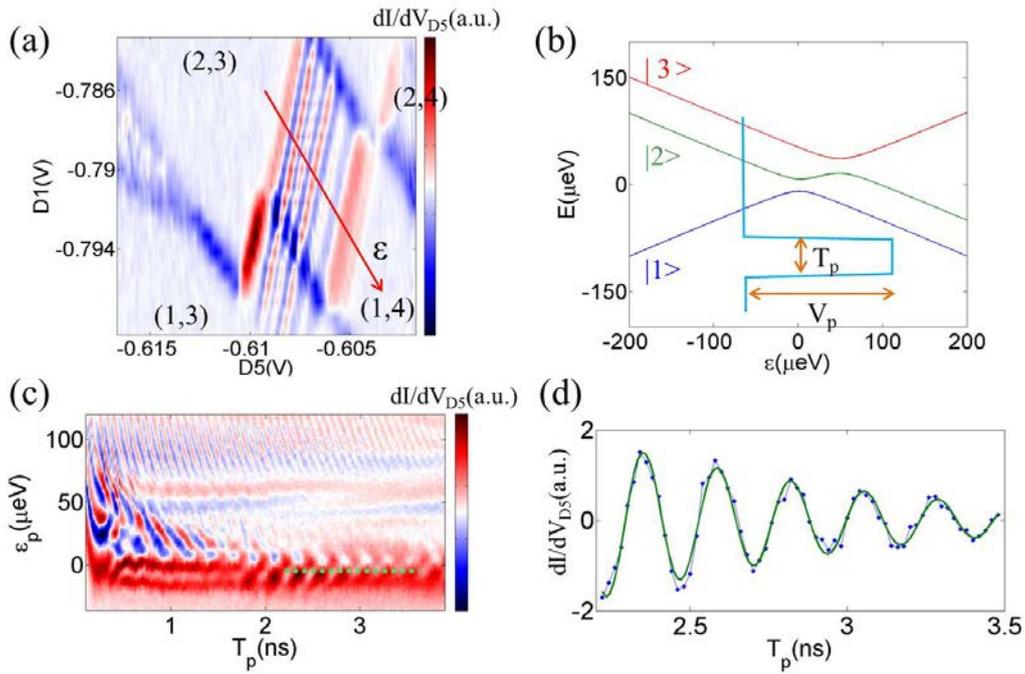

**FIG S1.** (a) QPC signal as a function of D1 and D5 voltages near the (2, 3) to (1, 4) charge transition. This is a blow-up of the corresponding transition in Fig. 1(b), with an additional short pulse of $T_p = 300$ ps, $V_p = 200$ μeV on top of the DC sweep to generate Larmor precessions. As a result, a series of interference fringes emerge right around, and parallel to, the charge transition line. (b) Calculated energy levels for the three lowest-energy states of the double quantum dot using the parameters: $2\Delta_1 = 4.2$ GHz, $2\Delta_2 = 5.0$ GHz, $\delta = 12.2$ GHz. (c) Coherent oscillations as a function of pulse width $T_p$ and detuning energy $\varepsilon_p$ with the pulse

amplitude $V_p$ fixed. Here $\varepsilon_p$ refers to the highest detuning the double dot can reach under the driving pulse. (d) A set of QPC signals as a function of the pulse width $T_p$. The blue line is a cut along the green dashed line in (c), while the green solid line is a best-fit curve with $T_2^*=2.12$ ns.